\documentclass[12pt]{article}

\usepackage{graphicx}
\usepackage{scicite}

\usepackage{times}

\newenvironment{sciabstract}{%
\begin{quote} \bf}
{\end{quote}}

\newcounter{lastnote}
\newenvironment{scilastnote}{%
\setcounter{lastnote}{\value{enumiv}}%
\addtocounter{lastnote}{+1}%
\begin{list}%
{\arabic{lastnote}.} {\setlength{\leftmargin}{.22in}}
{\setlength{\labelsep}{.5em}}} {\end{list}}

\title{Quantum Spin Hall Effect and Topological Phase Transition in HgTe Quantum Wells}

\author
{B. Andrei Bernevig,$^{1,2}$ Taylor L. Hughes,$^{1}$  and Shou-Cheng Zhang$^{1\ast}$\\
\\
\normalsize{$^{1}$Department of Physics, McCullough Building,
Stanford
University}\\
\normalsize{Stanford, CA 94305-4045}\\
\normalsize{$^{2}$Kavli Institute for Theoretical Physics}\\
\normalsize{University of California, Santa Barbara, CA 93106}
\\
\normalsize{$^\ast$To whom correspondence should be addressed
Shou-Cheng Zhang; E-mail:  sczhang@stanford.edu.} }

\date{}

\begin{document}


\baselineskip24pt


\maketitle


\begin{sciabstract}
We show that the Quantum Spin Hall Effect, a state of matter with
topological properties distinct from conventional insulators, can
be realized in HgTe/CdTe semiconductor quantum wells. By varying
the thickness of the quantum well, the electronic state changes
from a normal to an ``inverted" type at a critical thickness
$d_c$. We show that this transition is a topological quantum phase
transition between a conventional insulating phase and a phase
exhibiting the QSH effect with a single pair of helical edge
states. We also discuss the methods for experimental detection of
the QSH effect.
\end{sciabstract}


\newpage The spin Hall
effect\cite{d'yakonov1971,murakami2003,sinova2004,kato2004B,wunderlich2005}
has attracted great attention recently in condensed matter physics
both for its fundamental scientific importance, and its potentially
practical application in semiconductor spintronics. In particular,
the intrinsic spin Hall effect promises the possibility of designing
the intrinsic electronic properties of materials so that the effect
can be maximized. Based on this line of reasoning, it was
shown\cite{murakami2004A} that the intrinsic spin Hall effect can in
principle exist in band insulators, where the spin current can flow
without dissipation. Motivated by this suggestion, the quantum spin
Hall (QSH) effect has been proposed independently both for
graphene\cite{kane2005} and
semiconductors\cite{bernevig2006a,qi2005} , where the spin current
is carried entirely by the helical edge states in two dimensional
samples. Time reversal symmetry plays an important role in the
dynamics of the helical edge states\cite{kane2005A,wu2006,xu2006}.
When there are an even number of pairs of helical states at each
edge, impurity scattering or many-body interactions can open a gap
at the edge and render the system topologically trivial. However,
when there are an odd number of pairs of helical states at each edge
these effects cannot open a gap unless time reversal symmetry is
spontaneously broken at the edge. The stability of the helical edge
states has been confirmed in extensive numerical
calculations\cite{sheng2005,onoda2006}. The time reversal property
leads to the $Z_2$ classification\cite{kane2005A} of the QSH state.
States of matter can be classified according to their topological
properties. For example, the integer quantum Hall effect is
characterized by a topological integer $n$\cite{thouless1982},
which determines the quantized value of the Hall conductance and the
number of chiral edge states. It is invariant under smooth
distortions of the Hamiltonian, as long as the energy gap does not
collapse. Similarly, the number of helical edge states, defined
modulo two, of the QSH state is also invariant under topologically
smooth distortions of the Hamiltonian. Therefore, the QSH state is a
topologically distinct new state of matter, in the same sense as the
charge quantum Hall effect.

 Unfortunately, the initial proposal of the QSH in
graphene\cite{kane2005} was later shown to be
unrealistic\cite{yao2006,min2006}, as the gap opened by the
spin-orbit interaction turns out to be extremely small, of the order
of $10^{-3}$ meV. There are also no immediate experimental systems
available for the proposals in Ref.
\cite{bernevig2006a,murakami2006}. Here, we present theoretical
investigations of the type-III semiconductor quantum wells, and show
that the QSH state should be realized in the ``inverted" regime
where the well thickness $d$ is greater than a certain critical
thickness $d_c$. Based on general symmetry considerations and
standard $k \cdot p$ perturbation theory for
semiconductors\cite{kane1957}, we show that the electronic states
near the $\Gamma$ point are described by the relativistic Dirac
equation in $2+1$ dimensions. At the quantum phase transition at
$d=d_c$, the mass term in the Dirac equation changes sign, leading
to two distinct $U(1)$-spin and $Z_2$ topological numbers on either
side of the transition. Generally, knowledge of electronic states
near one point of the Brillouin Zone is insufficient to determine
the topology of the entire system, however, it does give robust and
reliable predictions on the change of topological quantum numbers.
The fortunate presence of a gap closing transition at the $\Gamma$
point in the HgTe/CdTe quantum wells therefore makes our theoretical
prediction of the QSH state conclusive.

The potential importance of inverted band gap semiconductors like
HgTe for the spin Hall effect was pointed out in
\cite{murakami2004A,qi2005}. The central feature of the type-III
quantum wells is band inversion: the barrier material such as CdTe
has a normal band progression, with the $\Gamma_6$ $s$-type band
lying above the $\Gamma_8$ $p$-type band, and the well material
HgTe having an inverted band progression whereby the $s$-type
$\Gamma_6$ band lies below the $p$-type $\Gamma_8$ band. In both
of these materials the gap is the smallest near the $\Gamma$ point
in the Brillouin zone (Fig. 1). In our discussion we neglect the
bulk split-off $\Gamma_7$ band, as it has negligible effects on
the band structure \cite{novik2005,pfeuffer-jeschke2000}.
Therefore, we shall restrict ourselves to a six band model, and
start with the following six basic atomic states per unit cell
combined into a six component spinor: \begin{equation}\Psi=\left(
|\Gamma_6,\frac{1}{2}\rangle,|\Gamma_6,-\frac{1}{2}\rangle,
|\Gamma_8,\frac{3}{2}\rangle,|\Gamma_8,\frac{1}{2}\rangle,|\Gamma_8,-\frac{1}{2}\rangle
|\Gamma_8,-\frac{3}{2}\rangle\right).
\nonumber\end{equation}\noindent In quantum wells grown in the
[001] direction the cubic, or spherical symmetry, is broken down
to the axial rotation symmetry in the plane. These six bands
combine to form the spin up and down ($\pm$) states of three
quantum well subbands: $E1,H1,L1$\cite{pfeuffer-jeschke2000}. The
$L1$ subband is separated from the other
two\cite{pfeuffer-jeschke2000}, and we neglect it, leaving an
effective four band model. At the $\Gamma$ point with in-plane
momentum $k_{\parallel} = 0$, $m_J$ is still a good quantum
number. At this point the $|E1, m_J\rangle$ quantum well subband
state is formed from the linear combination of the
$|\Gamma_6,m_J=\pm\frac{1}{2}\rangle$ and the
$|\Gamma_8,m_J=\pm\frac{1}{2}\rangle$ states, while the $|H1,
m_J\rangle$ quantum well subband state is formed from the
$|\Gamma_8,m_J=\pm\frac{3}{2}\rangle$ states. Away from the
$\Gamma$ point, the $E1$ and the $H1$ states can mix. As the
$|\Gamma_6,m_J=\pm\frac{1}{2}\rangle$ state has even parity, while
the $|\Gamma_8,m_J=\pm\frac{3}{2}\rangle$ state has odd parity
under two dimensional spatial reflection, the coupling matrix
element between these two states must be an odd function of the
in-plane momentum $k$. From these symmetry considerations, we
deduce the general form of the effective Hamiltonian for the $E1$
and the $H1$ states, expressed in the basis of ${|E1,
m_J=1/2\rangle, |H1, m_J=3/2\rangle}$ and ${|E1, m_J=-1/2\rangle,
|H1, m_J=-3/2\rangle}$:
\begin{equation}
H_{eff}(k_x,k_y)=\left(\matrix{ H(k)&0 \cr 0&H^*(-k)} \right), \ \
H=\epsilon(k) + d_i(k) \sigma_i
 \label{dirac}
\end{equation}\noindent
where $\sigma_i$ are the Pauli matrices. The form of $H^*(-k)$ in
the lower block is determined from time reversal symmetry and $H^*
(-k)$ is unitarily equivalent to $H^*(k)$ for this system(see
Supporting Online Material). If inversion symmetry and axial
symmetry around the growth axis are not broken then the
inter-block matrix elements vanish, as presented.

We see that, to the lowest order in $k$, the Hamiltonian matrix
decomposes into $2\times 2$ blocks. From the symmetry arguments
given above, we deduce that $d_3(k)$ is an even function of $k$,
while $d_1(k)$ and $d_2(k)$ are odd functions of $k$. Therefore, we
can generally expand them in the following form:
\begin{eqnarray}
& d_1+id_2=A(k_x+i k_y)\equiv Ak_+ \nonumber \\
& d_3 = M-B (k_{x}^2+k_{y}^2) \ , \ \epsilon_k =
C-D(k_{x}^2+k_{y}^2). \label{d-expansion}
\end{eqnarray}\noindent
The Hamiltonian in the $2\times 2$ subspace therefore takes the
form of the $2+1$ dimensional Dirac Hamiltonian, plus an
$\epsilon(k)$ term which drops out in the quantum Hall response.
The most important quantity is the mass, or gap parameter $M$,
which is the energy difference between the $E1$ and $H1$ levels at
the $\Gamma$ point. The overall constant $C$ sets the zero of
energy to be the top of the valence band of bulk HgTe. In a
quantum well geometry, the band inversion in HgTe necessarily
leads to a level crossing at some critical thickness $d_c$ of the
HgTe layer. For thickness $d<d_c$, i.e. for a thin HgTe layer, the
quantum well is in the ``normal" regime, where the CdTe is
predominant and hence the band energies at the $\Gamma$ point
satisfy $E(\Gamma_6)> E(\Gamma_8)$. For $d> d_c$ the HgTe layer is
thick and the well is in the inverted regime where HgTe dominates
and $E(\Gamma_6)< E(\Gamma_8)$. As we vary the thickness of the
well, the $E1$ and $H1$ bands must therefore cross at some $d_c$,
and the gap parameter $M$ changes sign between the two sides of
the transition (Fig. 2). Detailed calculations show that, close to
transition point, the $E1$ and $H1$ band, both doubly degenerate
in their spin quantum number, are far away in energy from any
other bands\cite{pfeuffer-jeschke2000}, hence making an effective
Hamiltonian description possible. In fact, the form of the
effective Dirac Hamiltonian and the sign change of $M$ at $d=d_c$
for the HgTe/CdTe quantum wells deduced above from general
arguments is already completely sufficient to conclude the
existence of the QSH state in this system. For the sake of
completeness, we also provide the microscopic derivation directly
from the Kane model using realistic material parameters (see SOM).

Fig. 2 shows the energies of both the $E1$ and $H1$ bands at
$k_{\parallel}=0$ as a function of quantum well thickness $d$
obtained from our analytical solutions. At $d = d_c \sim 64 \AA$
these bands cross. Our analytic results are in excellent
qualitative and quantitative agreement with previous numerical
calculations for the band structure of
Hg$_{1-x}$Cd$_x$Te/HgTe/Hg$_{1-x}$Cd$_x$Te quantum wells
\cite{pfeuffer-jeschke2000,novik2005}. We also observe that for
quantum wells of thickness $40 \AA < d < 70\AA ,$ close to $d_c$,
the $E1 \pm$ and $H1\pm$ bands are separated from all other bands
by more than $30$ meV\cite{pfeuffer-jeschke2000}.

Let us now define an ordered set of four $6$-component basis
vectors $ \psi_{1,...,4} = (|E1,+>,|H1,+>,|E1,->,|H1,->)$ and
obtain the Hamiltonian at non-zero in-plane momentum in
perturbation theory. We can write the effective $4\times 4$
Hamiltonian for the $E1\pm, \; H1\pm$ bands as:
\begin{equation}
H^{eff}_{ij}(k_x,k_y)=\int_{-\infty}^{\infty} dz
<\psi_j|{\cal{H}}(k_x, k_y, -i \partial_z)|\psi_i>.
\end{equation}
\noindent The form of the effective Hamiltonian is severely
constrained by symmetry with respect to $z$. Each band has a
definite $z$ symmetry or antisymmetry and vanishing matrix
elements between them can be easily identified. For example,
$H^{eff}_{23} = \frac{1}{\sqrt{6}} P k_+ \int_{-\infty}^{\infty}
dz \langle \Gamma_6, +\frac{1}{2} (z) | \Gamma_8, -\frac{1}{2} (z)
\rangle $ vanishes because $|\Gamma_6, +\frac{1}{2} \rangle (z)$
is even in $z$ whereas $| \Gamma_8, -\frac{1}{2}  \rangle (z)$ is
odd. The procedure yields exactly the form of the effective
Hamiltonian (\ref{dirac}) as we anticipated from the general
symmetry arguments, with the coupling functions taking exactly the
form of (\ref{d-expansion}). The dispersion relations (see SOM)
have been checked to be in agreement with prior numerical results
\cite{pfeuffer-jeschke2000,novik2005}. We note that for $k \in [0,
0.01 {\AA^{-1}}]$ the dispersion relation is dominated by the
Dirac linear terms. The numerical values for the coefficients
depend on the thickness, and for values at $d=58 \AA$  and
$d=70\AA\;$ see SOM.

Having presented the realistic $k \cdot p$ calculation starting
from the microscopic 6-band Kane model, we now introduce a
simplified tight binding model for the $E1$ and the $H1$ states
based on their symmetry properties. We consider a square lattice
with four states per unit cell. The $E1$ states are described by
the $s$-orbital states $\psi_{1,3}=|s,\alpha=\pm 1/2\rangle$, and
the $H1$ states are described by the spin-orbit coupled
$p$-orbital states $ \psi_{2,4}=\pm \frac{1}{\sqrt{2}}|p_x\pm i
p_y,\alpha=\pm 1/2\rangle$. Here $\alpha$ denotes the electron
spin. Nearest neighbor coupling between these states gives the
tight-binding Hamiltonian of the form of Eq. \ref{dirac}, with the
matrix elements given by
\begin{eqnarray}
& d_1+id_2=A(\sin(k_x)+i \sin(k_y)) \nonumber \\
& d_3 = - 2B( 2 - \frac{M}{2B} - \cos(k_x) -\cos(k_y)) \nonumber \\
& \epsilon_k = C-2D(2 - \cos(k_x) - \cos(k_y)).
\label{tight-binding}
\end{eqnarray}\noindent
The tight-binding lattice model simply reduces to the continuum
model Eq. \ref{dirac} when expanded around the $\Gamma$ point. The
tight-binding calculation serves dual purposes. For readers
uninitiated in the Kane model and $k \cdot p$ theory, this gives a
simple and intuitive derivation of our effective Hamiltonian that
captures all the essential symmetries and topology. On the other
hand, it also introduces a short-distance cut-off so that the
topological quantities can be well-defined.

Within each $2 \times 2$ sub-block, the Hamiltonian is of the
general form studied in Ref. \cite{qi2005}, in the context of the
quantum anomalous Hall effect, where the Hall conductance is given
by:
\begin{eqnarray}
\sigma_{xy}&=&-\frac{1}{8\pi^2}\int\int dk_xdk_y{\bf
\hat{d}}\cdot{\bf
\partial_x\hat{d}\times\partial_y\hat{d}}\label{conductQAHE}
\end{eqnarray}
\noindent in units of $e^2/h$, where ${\bf \hat{d}}$ denotes the
unit $d_i(k)$ vector introduced in the Hamiltonian Eq.
\ref{dirac}. When integrated over the full Brillouin Zone,
$\sigma_{xy}$ is quantized to take integer values which measures
the Skyrmion number, or the number of times the unit ${\bf
\hat{d}}$ winds around the unit sphere over the Brillouin Zone
torus. The topological structure can be best visualized by
plotting ${\bf \hat{d}}$ as a function of $k$. In a Skyrmion with
a unit of topological charge, the ${\bf \hat{d}}$ vector points to
the north (or the south) pole at the origin, to the south (or the
north) pole at the zone boundary, and winds around the equatorial
plane in the middle region.

Substituting the continuum expression for the $d_i(k)$ vector as
given in Eq. \ref{d-expansion}, and cutting off the integral at
some finite point in momentum space, one obtains $\sigma_{xy} =
\frac{1}{2} sign (M)$, which is a well-known result in field
theory\cite{redlich1984}. In the continuum model, the ${\bf
\hat{d}}$ vector takes the configuration of a meron, or half of a
Skyrmion, where it points to the north (or the south) pole at the
origin, and winds around the equator at the boundary. As the meron
is half of a Skyrmion, the integral Eq. \ref{conductQAHE} gives
$\pm \frac{1}{2}$. The meron configuration of the $d_i(k)$ is
depicted in Fig. 2. In a non-interacting system, half-integral
Hall conductance is not possible, which means that other points
from the Brillouin Zone must either cancel or add to this
contribution so that the total Hall conductance becomes an
integer. The fermion-doubled partner\cite{NIELSEN1981} of our
low-energy fermion near the $\Gamma$-point lies in the higher
energy spectrum of the lattice and contributes to the total
$\sigma_{xy}.$ Therefore, our effective Hamiltonian near the
$\Gamma$ point can not give a precise determination of the Hall
conductance for the whole system. However, as one changes the
quantum well thickness $d$ across $d_c$, $M$ changes sign, and the
gap closes at the $\Gamma$ point leading to a vanishing $d_i(k=0)$
vector at the transition point $d=d_c$. The sign change of $M$
leads to a well-defined change of the Hall conductance $\Delta
\sigma_{xy} = 1$ across the transition. As the $d_i(k)$ vector is
regular at the other parts of the Brillouin Zone, they can not
lead to any discontinuous changes across the transition point at
$d=d_c$. So far, we have only discussed one $2\times 2$ block of
the effective Hamiltonian $H$. General time reversal symmetry
dictates that $\sigma_{xy}(H)=-\sigma_{xy}(H^*)$, therefore, the
total charge Hall conductance vanishes, while the spin Hall
conductance, given by the difference between the two blocks, is
finite, and given by $\Delta \sigma^{(s)}_{xy} = 2$. From the
general relationship between the quantized Hall conductance and
the number of edge states\cite{qi2006}, we conclude that the two
sides of the phase transition at $d=d_c$ must differ in the number
of pairs of helical edge states by one, thus concluding our proof
that one side of the transition must be $Z_2$ odd, and
topologically distinct from a fully gapped conventional insulator.

It is desirable to establish which side of the transition is
indeed topologically non-trivial. For this purpose, we return to
the tight-binding model Eq. \ref{tight-binding}. The Hall
conductance of this model has been calculated \cite{qi2006} in the
context of the quantum anomalous Hall effect, and previously in
the context of lattice fermion simulation\cite{golterman1993}.
Besides the $\Gamma$ point, which becomes gapless at $M/2B=0$,
there are three other high symmetry points in the Brillouin Zone.
The $(0,\pi)$ and the $(\pi, 0)$ points become gapless at
$M/2B=2$, while the $(\pi,\pi)$ point becomes gapless at $M/2B=4$.
Therefore, at $M/2B=0$, there is only one gapless Dirac point per
$2\times 2$ block. This behavior is qualitatively different from
the Haldane model of graphene\cite{haldane1988}, which has two
gapless Dirac points in the Brillouin Zone. For $M/2B<0$,
$\sigma_{xy}=0$, while $\sigma_{xy}=1$ for $0<M/2B<2$. As this
condition is satisfied in the inverted gap regime where $M/2B
=2.02 \times 10^{-4} $ at $70 \AA\;$ (see SOM), and not in the
normal regime (where $M/2B < 0$), we believe that the inverted
case is the topologically non-trivial regime supporting a QSH
state.

We now discuss the experimental detection of the QSH state. A
series of purely electrical measurements can be used to detect the
basic signature of the QSH state. By sweeping the gate voltage,
one can measure the two terminal conductance $G_{LR}$ from the
p-doped to bulk-insulating to n-doped regime(Fig. 3). In the bulk
insulating regime, $G_{LR}$ should vanish at low temperatures for
a normal insulator at $d<d_c$, while $G_{LR}$ should approach a
value close to $2e^2/h$ for $d>d_c$. Strikingly, in a six terminal
measurement, the QSH state would exhibit vanishing electric
voltage drop between the terminals $\mu_1$ and $\mu_2$ and between
$\mu_3$ and $\mu_4$, in the zero temperature limit and in the
presence of a finite electric current between the $L$ and $R$
terminals. In other words, longitudinal resistance should vanish
in the zero temperature limit with a power law dependence, over
distances larger than the mean free path. Because of the absence
of backscattering, and before spontaneous breaking of time
reversal sets in, the helical edge currents flow without
dissipation, and the voltage drop occurs only at the drain side of
the contact\cite{wu2006}. The vanishing of the longitudinal
resistance is one of the most remarkable manifestations of the QSH
state. Finally, a spin filtered measurement can be used to
determine the spin-Hall conductance $\sigma^{(s)}_{xy}$. Numerical
calculations\cite{sheng2005} show that it should take a value
close to $\sigma^{(s)}_{xy}=2\frac{e^2}{h}$.

Constant experimental progress on HgTe over the past two decades
makes the experimental realization of our proposal possible. The
mobility of the HgTe/CdTe quantum wells has reached $\mu \sim
6\times 10^ 5 cm^2/(V s)$\cite{Ortner2000}. Experiments have
already confirmed the different characters of the upper band below
($E1$) and above ($H1$) the critical thickness $d_c$
\cite{novik2005,BECKER2000}. The experimental results are in
excellent agreement with band-structure calculations based on the
$k \cdot p$ theory. Our proposed two terminal and six terminal
electric measurements can be carried out on existing samples
without radical modification, with samples of $d<d_c\approx 64\AA$
and $d>d_c\approx 64\AA$ yielding contrasting results. Following
this detailed proposal, we believe that the experimental detection
of the QSH state in HgTe/CdTe quantum wells is possible.

\newpage
\bibliography{HgTe}

\bibliographystyle{Science}

\begin{scilastnote}
\item We wish to thank X. Dai, Z. Fang, F.D.M. Haldane, A. H.
MacDonald, L. W. Molenkamp, N. Nagaosa, X.-L. Qi, R. Roy, and R.
Winkler for discussions. B.A.B. wishes to acknowledge the
hospitality of the Kavli Institute for Theoretical Physics at
University of California at Santa Barbara, where part of this work
was performed. This work is supported by the NSF through the
grants DMR-0342832, and by the US Department of Energy, Office of
Basic Energy Sciences under contract DE-AC03-76SF00515.

\end{scilastnote}
\newpage
\section{Figures }
\begin{figure*}[h]
        \includegraphics[scale=0.50]{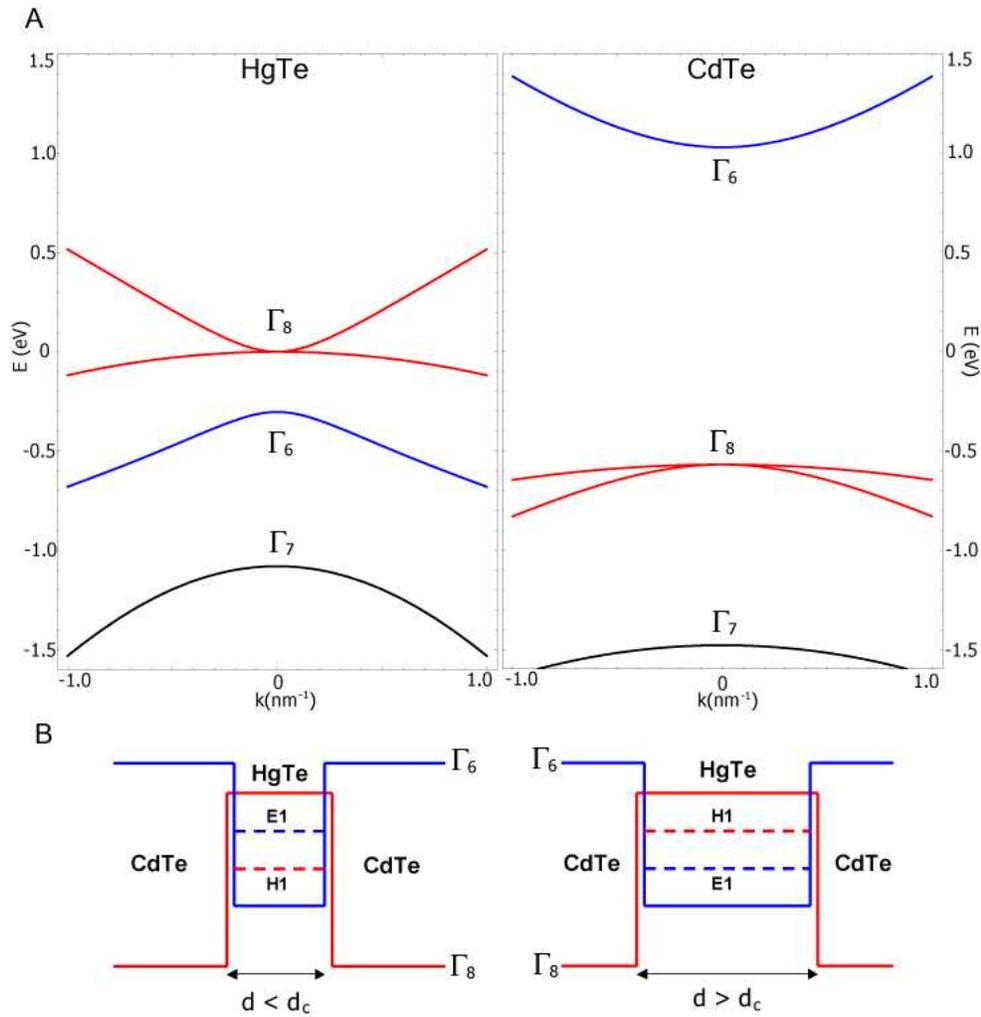}
\caption{ (A) Bulk energy bands of HgTe and CdTe near the $\Gamma$
point. (B) The CdTe/HgTe/CdTe quantum well in the normal regime
$E1> H1$ with $d< d_c$ and in the inverted regime $H1> E1$ with
$d> d_c$. In this, and all subsequent figures $\Gamma_8$/H1
($\Gamma_6$/E1) symmetry is correlated with the color red
(blue).}\label{fig1}
  \end{figure*}

\begin{figure*}
        \includegraphics[scale=0.50]{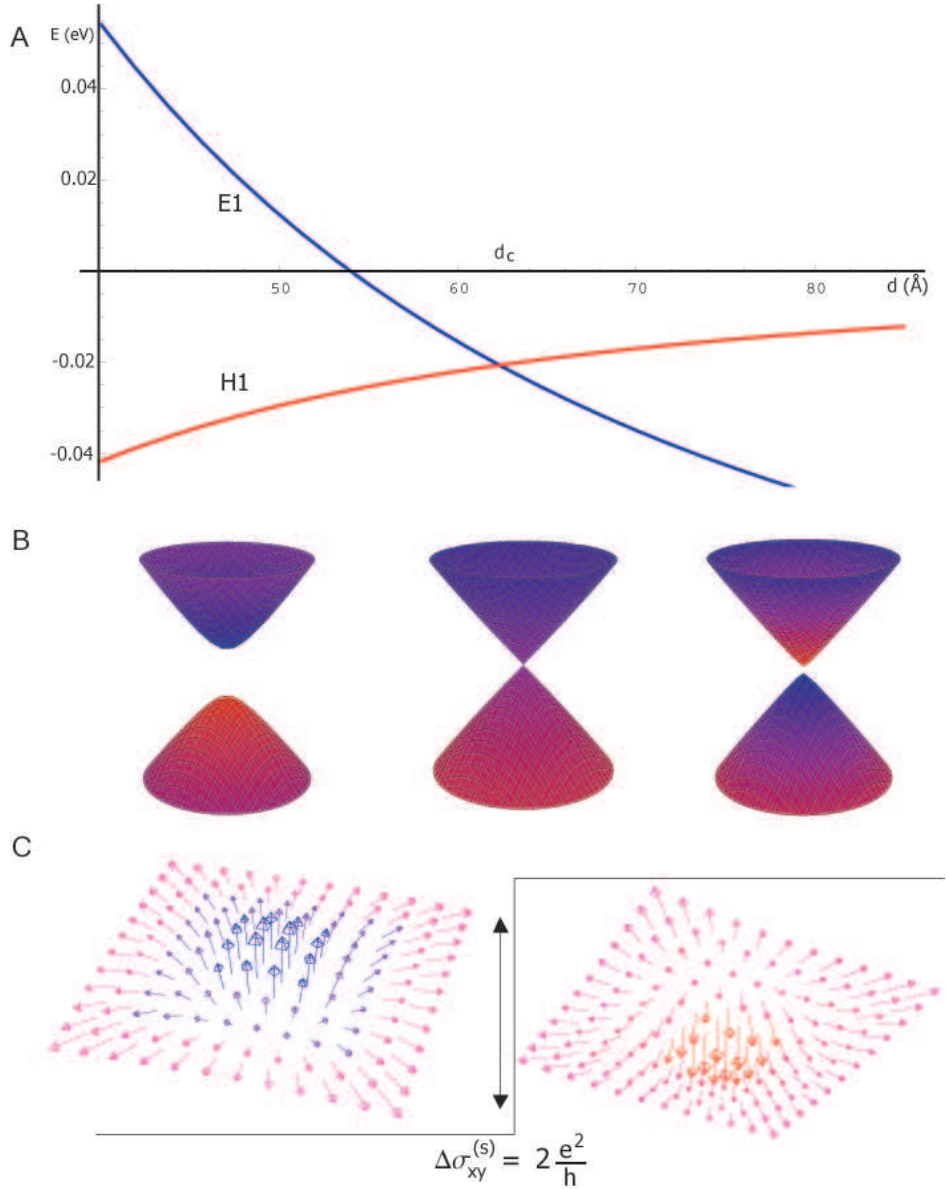}
\caption{(A) Energy (eV) of $E1$ (blue) and $H1$ (red) bands at
$k_{\parallel}=0$ vs. quantum-well thickness $d$ ($\AA$). (B)
Energy dispersion relations $E(k_x,k_y)$ of the $E1,H1$ subbands
at $40 \AA,$ $63.5 \AA$ and $70 \AA$ from left to right. Colored
shading indicates the symmetry type of band at that $k$-point.
Places where the cones are more red (blue) indicates that the
dominant states are H1 (E1) states at that point. Purple shading
is a region where the states are more evenly mixed. For $40\AA$
the lower (upper) band is dominantly H1(E1). At $63.5\AA$ the
bands are evenly mixed near the band crossing and retain their
$d<d_c$ behavior moving further out in $k$-space. At $d=70\AA$ the
regions near $k_{\parallel}=0$ have flipped their character but
eventually revert back to the $d<d_c$ further out in $k$-space.
Only this dispersion shows the meron structure (red and blue in
the same band). (C) Schematic meron configurations representing
the $d_i(k)$ vector near the $\Gamma$ point. The shading of the
merons has the same meaning as the dispersion relations above. The
change in meron number across the transition is exactly equal to
$1$, leading to a quantum jump of the spin Hall conductance
$\Delta\sigma_{xy}^{(s)}=2e^2/h.$ We measure all Hall conductances
in electrical units. All of these plots are for
Hg$_{0.32}$Cd$_{0.68}$Te/HgTe quantum wells.
 }\label{fig2}
  \end{figure*}

\begin{figure*}
        \includegraphics[scale=0.75]{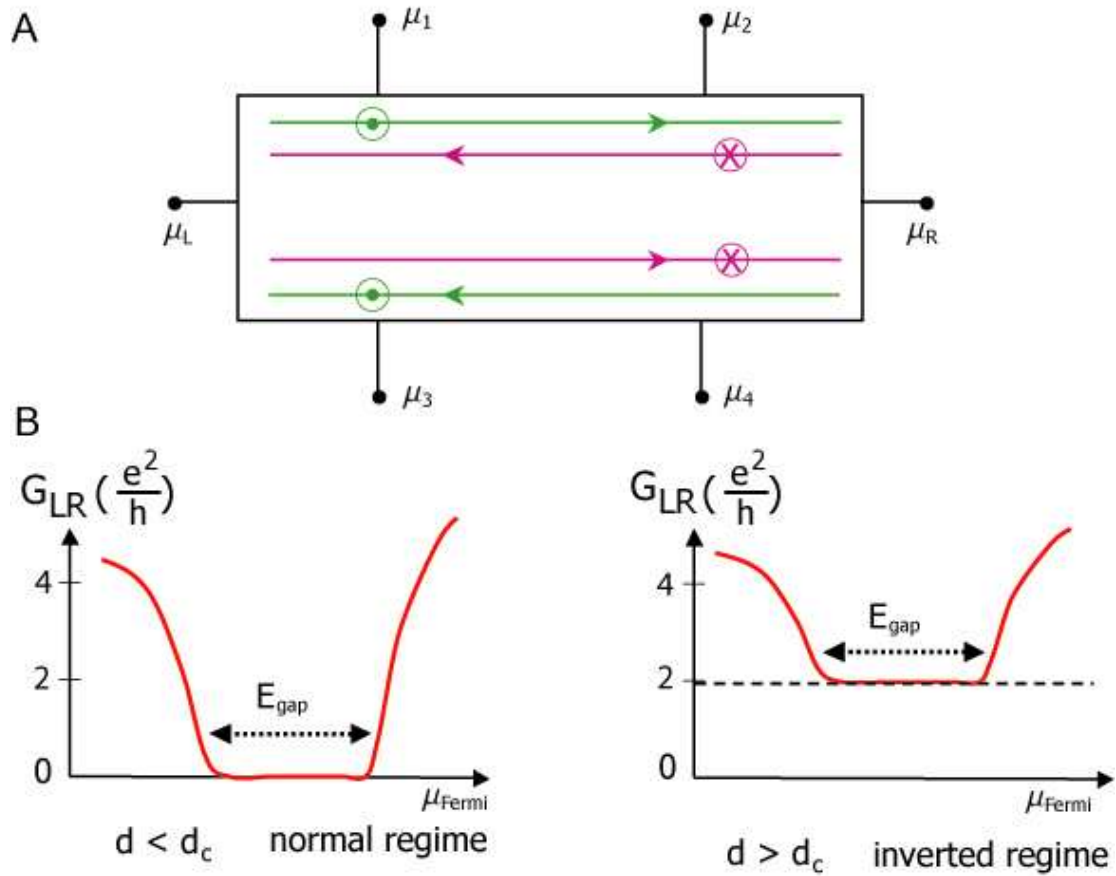}
\caption{(A) Experimental setup on a six terminal Hall bar showing
pairs of edge states with spin up (down) states green (purple).
(B)A two-terminal measurement on a Hall bar would give $G_{LR}$
close to $2e^2/h$ contact conductance on the QSH side of the
transition and zero on the insulating side. In a six-terminal
measurement, the longitudinal voltage drops $\mu_2 -\mu_1$ and
$\mu_4 - \mu_3$ vanish on the QSH side with a power law as the
zero temperature limit is approached. The spin-Hall conductance
$\sigma_{xy}^{(s)}$ has a plateau with the value close to $2
\frac{e^2}{h}.$ }\label{fig3}
  \end{figure*}

\newpage
\section{Supporting Online Material}
We show that our effective Hamiltonian can be derived
perturbatively with $\textbf{k}\cdot\textbf{P}$ theory and is a
quantitatively accurate description of the band structure for the
$E1$ and the $H1$ subband states. We start from the $6$-band bulk
Kane model which incorporates the $\Gamma_6$ and $\Gamma_8$ bands
but neglects the split-off $\Gamma_7$ band( the contribution of
the split-off band to the E1,H1 energies is less than $5\%$({\it
S2})):
\begin{equation}
{\cal{H}}(\vec{k}) = \left(%
\matrix{
  E_c I_{2\times 2}+ H_c & T_{2\times 4} \cr
  T^\dagger_{4 \times 2} & E_v I_{4\times 4} + H_v \\
}%
\right) \label{kanehamiltonian}
\end{equation}
\noindent where $E_c$ is the conduction band offset energy, $E_v$
is the valence band offset energy, $H_c, H_v$ are the conduction
and valence (Luttinger) band Hamiltonians, while $T(k)$ is the
interaction matrix between the conduction and valence bands:
\begin{eqnarray}
& H_c = \left(%
\matrix{
  \frac{\hbar^2 k^2}{2m^\star} & 0 \cr
  0 & \frac{\hbar^2 k^2}{2m^\star} \cr
}%
\right); \;\; T^\dagger = \left(%
\matrix{
  -\frac{1}{\sqrt{2}} P k_- & 0 \cr
  \sqrt{\frac{2}{3}} P k_z & -\frac{1}{\sqrt{6}} P k_- \cr
  \frac{1}{\sqrt{6}} P k_-  & \sqrt{\frac{2}{3}} P k_z \cr
  0 & \frac{1}{\sqrt{2}} P k_- \cr
}%
\right) \nonumber \\ & H_v  = - \frac{\hbar^2}{2m_0} (\gamma_1+
\frac{5}{2} \gamma_2) k^2 + \frac{\hbar^2}{m_0}\gamma_2 (\vec{k}
\cdot \vec{S})^2
\end{eqnarray}
\noindent where $m^\star$ is the effective electron mass in the
conduction band,  $k_\pm = k_x \pm i k_y$, $P = -\frac{\hbar}{m_0}
\langle s | p_x | X \rangle$ is the Kane matrix element between
the $s$ and $p$ bands with $m_0$ the bare electron mass, $\vec{S}$
is the spin-$3/2$ operator whose representation are the $4\times
4$ spin matrices, and $\gamma_1, \gamma_2$ are the effective
Luttinger parameters in the valence band. The ordered basis for
this form of the Kane model is
$(|\Gamma^6,+1/2>,|\Gamma^6,-1/2>,|\Gamma^8,+3/2>,|\Gamma^8,+1/2>,|\Gamma^8,-1/2>,|\Gamma^8,-3/2>).$
Although due to space constraints the above Hamiltonian is written
in the spherical approximation, our calculations include the
anisotropy effects generated by a third Luttinger parameter
$\gamma_3 \ne \gamma_2$. The quantum well growth direction is
along $z$  with Hg$_{1-x}$Cd$_x$Te for $z< -d/2$, HgTe for $-d/2 <
z < d/2$ and Hg$_{1-x}$Cd$_x$Te for $z> d/2$. Our problem reduces
to solving, in the presence of continuous boundary conditions, the
Hamiltonian Eq. \ref{kanehamiltonian} in each of the $3$ regions
of the quantum well. The material parameters $E_c, E_v,
\gamma_{1,2,3}$, and $m^\star$, are discontinuous at the
boundaries of the $3$ regions, taking the
Hg$_{1-x}$Cd$_x$Te/HgTe/Hg$_{1-x}$Cd$_x$Te values (see Ref. ({\it
S1}) for numerical values), so we have a set of $3$
multi-component eigenvalue equations coupled by the boundary
conditions. The in-plane momentum is a good quantum number and the
solution in each region takes the general form:
\begin{eqnarray} & {\cal{H}}(\vec{k}) \psi(k_x, k_y, z) = {\cal{H}}(k_x,
k_y, -i \partial_z)\psi(k_x, k_y, z) ; \nonumber
\\
& \psi (k_x,k_y, z) = e^{i(k_x x + k_y y)}\Psi(z)
\end{eqnarray}
\noindent where $\Psi (z)$ is the envelope function spinor in the
six component basis introduced earlier.

 We solve the Hamiltonian analytically by first solving for the
eigenstates at zero in-plane momentum, and then perturbatively
finding the form of the Hamiltonian for finite in-plane
$\textbf{k}$: ${\cal{H}}(k_x, k_y, -i
\partial_z) = {\cal{H}}(0, 0, -i \partial_z) + \delta
{\cal{H}}(k_x, k_y, -i
\partial_z)$. At $k_x=k_y=0$ we have the following Hamiltonian:
\begin{equation}
{\cal{H}}(0,0,-i\partial_z)=\left(\matrix{
T&0&0&\sqrt{\frac{2}{3}}P (-i \partial_z) &0&0\cr
0&T&0&0&\sqrt{\frac{2}{3}}P (-i \partial_z) &0\cr 0&0&U+V&0&0&0\cr
\sqrt{\frac{2}{3}}P (-i \partial_z)&0&0&U-V&0&0\cr
0&\sqrt{\frac{2}{3}}P (-i \partial_z)&0&0&U-V&0\cr
0&0&0&0&0&U+V}\right)
\end{equation}
\noindent where $T= E_c(z) + (-\partial_{z} A(z)\partial_{z})$,
$U=E_v(z) - (-\partial_{z} \gamma_1(z) \partial_{z})$,
$V=2(-\partial_z \gamma_2(z) \partial_z).$  These parameters are
treated as step functions in the $z$-direction with an abrupt
change from the barrier region to the well region.

A general state in the envelope function approximation can be
written in the following form:
\begin{equation}
\Psi(k_x,k_y,z)=e^{i(k_x x+k_y y)}\left( \matrix{ f_1(z)\cr
f_2(z)\cr f_3(z)\cr f_4(z)\cr f_5(z)\cr
f_6(z)}\right).\end{equation} \noindent At $k_x =k_y =0$ the $f_3$
and $f_6$ components decouple and form the spin up and down
($\pm$) states of the $H1$ subband. The $f_1,f_2,f_4,f_5$
components combine together to form the spin up and down ($\pm$)
states of the E1 and L1 subbands. The linear-in-$k_z$ operator
$\sqrt{\frac{2}{3}} P k_z,$  in ${\cal{H}}(0, 0, -i
\partial_z)$ forces the  $|\Gamma_6, \pm \frac{1}{2} \rangle (z) $ and
$|\Gamma_8, \pm \frac{1}{2} \rangle (z)$ components of the $E1$
band to have different reflection symmetry under $z\leftrightarrow
-z$. The $|\Gamma_6\rangle$ band is symmetric in $z$
(exponentially decaying in CdTe and a $\cosh(z)$ dependence in
HgTe) while the $|\Gamma_8, m_J=\pm 1/2\rangle$ band is
antisymmetric in $z$ (exponentially decaying in CdTe and a
$\sinh(z)$ dependence in HgTe). The opposite choice of symmetry
under $z\rightarrow -z$ reflection leads to the $L1$ band.
However, this band is far away in energy from both the $E1$ and
the $H1$ bands, does not cross either of them in the region of
interest({\it S2}), and we hence discard it.

For the E1 band we take the ansatz, already knowing it must be an
interface state({\it S2}), to be:
\begin{eqnarray}
\Psi_I = \left( \matrix{ e^{\alpha z} C_1 \cr 0 \cr\ 0\cr
e^{\alpha z} C_4\cr 0\cr 0}\right), & \Psi_{II} = \left( \matrix{(
e^{\delta z}+e^{-\delta z}) V_1 \cr 0 \cr 0\cr (e^{\delta
z}-e^{-\delta z}) V_4\cr 0\cr 0}\right), & \Psi_{III} = \left(
\matrix{ e^{-\alpha z} C_1 \cr 0 \cr 0\cr -e^{-\alpha z} C_4\cr
0\cr 0}\right).\end{eqnarray} \noindent If we act on this ansatz
with the Hamiltonian we decouple the $6\times 6$ matrix into two,
coupled, one-dimensional Schrodinger equations:\begin{eqnarray} T
f_1(z)
+\sqrt{\frac{2}{3}}P(z) (-i \partial_z) f_4(z) = E f_1(z) \\
\sqrt{\frac{2}{3}}P(z) (-i \partial_z)f_1(z) +(U-V)f_4(z)=E
f_4(z)\end{eqnarray} \noindent where $P,T,U,V$ are given above.
Using the restrictions from the Hamiltonian, the continuity of
each wavefunction component at the boundaries, and the continuity
of the probability current across the boundary, we derive the
following set of equations that determine $\alpha$ and $\delta$ as
a function of E:
\begin{eqnarray}
\frac{E_{c}^{(Cd)}-A^{(Cd)}\alpha^2 (E)
-E}{\sqrt{\frac{2}{3}}\frac{P}{i}
\alpha(E)}=\frac{\sqrt{\frac{2}{3}}\frac{P}{i}
\alpha(E)}{E_{v}^{(Cd)}+(\gamma_{1}^{(Cd)}+2\gamma_{2}^{(Cd)})\alpha^2
(E) -E}\\ \frac{E_{c}^{(Hg)}-A^{(Hg)}\delta^2 (E)
-E}{\sqrt{\frac{2}{3}}\frac{P}{i}
\delta(E)}=\frac{\sqrt{\frac{2}{3}}\frac{P}{i}
\delta(E)}{E_{v}^{(Hg)}+(\gamma_{1}^{(Hg)}+2\gamma_{2}^{(Hg)})\delta^2
(E) -E}\end{eqnarray} \noindent where a parameter $X^{(Cd)}$ means
the value of that parameter in the CdTe barrier material and
$X^{(Hg)}$ is the value in the HgTe well material. Once we have
$\alpha(E)$ and $\delta(E)$ we can use them to determine $E$
through the following equation derived from the boundary
conditions:
\begin{equation}
\frac{E_{c}^{(Cd)}-A^{(Cd)}\alpha^2(E)-E}{\alpha(E)}=-
\tanh\left(\frac{\delta(E)
d}{2}\right)\left(\frac{E_{c}^{(Hg)}-A^{(Hg)}\delta^2(E)-E}{\delta(E)}\right).\end{equation}\noindent
These rational transcendental equations are solved numerically to
obtain the energy of the $E1$ subband at $k_x=k_y=0.$

We can follow a similar procedure to derive the energy of the $H1$
subband. The heavy hole subband (at $k_x=k_y=0$) completely
decouples from the other bands and we have the one-dimensional,
one-component Hamiltonian:
\begin{equation} H f_3(z)= E_v(z)-(\gamma_1(z)
-2\gamma_2(z))(-\partial_{z}^{2})f_3(z)= E f_{3}(z).\end{equation}
\noindent We have the wavefunction in three regions
\begin{equation}
\left( \matrix{ \Psi_{I}(z)\cr \Psi_{II}(z) \cr
\Psi_{III}(z)}\right) = \left( \matrix{ C_3 e^{\beta z}\cr V_3
\cos(\kappa z) \cr C_3 e^{-\beta z}}\right)\end{equation}\noindent
where $\beta^2
(E)=\frac{E-E_{v}^{(Cd)}}{\gamma_{1}^{(Cd)}-2\gamma_{2}^{(Cd)}}$
and $\kappa^2
(E)=\frac{E_{v}^{(Hg)}-E}{\gamma_{1}^{(Hg)}-2\gamma_{2}^{(Hg)}}.$
We can pick $C_3=1$ which gives us the relation \begin{equation}
V_3=\frac{e^{-\frac{1}{2}\beta d}}{\cos(\kappa(E)
d/2)}\end{equation}\noindent from the boundary condition at
$z=-d/2.$ Finally, we need to normalize the wavefunction to get
the coefficients. The energy of this state is determined by
considering the conservation of probability current across the
boundary. The following equation is solved for the energy:
\begin{equation}
\frac{1}{(\gamma_{1}^{(Cd)}-2\gamma_{2}^{(Cd)})\beta(E)}=\frac{1}{(\gamma_{1}^{(Hg)}-2\gamma_{2}^{(Hg)})\kappa
(E)}\cot(\kappa(E) d/2).\end{equation}

We repeat both of these processes on a state with only $f_{2}(z)$
and $f_{5}(z)$ non-zero and on a state with only $f_{6}(z)$
non-zero and to get the $E1^{-}$ and $H1^{-}$ bands respectively.
We have the forms of these states at $k_x = k_y =0$ and can use $k
\cdot P$ perturbation theory to derive a two dimensional
Hamiltonian near the $\Gamma$ point in $k$-space.

\section{Perturbation Theory and Effective Hamiltonian}
Define an ordered set of basis vectors $
(|E1,+>,|H1,+>,|E1,->,|H1,->).$ We can write the effective
Hamiltonian as:
\begin{equation}
H_{ij}(k_x,k_y)=\int_{-\infty}^{\infty} dz <\psi_j|H_{6\times
6}(k_x,k_y,-i\partial_z)|\psi_i>\end{equation} \noindent where
$\psi_i$ is the $i$-th element of the basis set given above which
will give a $4\times 4$ effective Hamiltonian. The integrals must
be split into the three regions defined above, and the parameters
from each material must be accounted for in the Hamiltonian. This
Hamiltonian depends on the quantum-well width $d,$ and once $d$ is
specified we can numerically calculate the matrix-elements. It is
important to note that $f_1 (z),f_2 (z),f_3 (z),f_6 (z)$ are
symmetric with respect to $z$ and $f_4 (z),f_5 (z)$ are
antisymmetric in $z$({\it S2}) which is a useful simplification in
performing the integrals. An example of one integral is of the
form:\begin{eqnarray} \int dz f_{3}^{*}(z)
\left\{\gamma_3(z),-i\partial_z\right\}f_{4}(z)= \nonumber
\\ \int dz \frac{1}{i}\left(2\gamma_{3}(z) f_{3}^{*}(z)\partial_z f_{4}(z)+f_{3}^{*}(z)f_{4}(z)\partial_z \gamma_3
(z)\right).\end{eqnarray}The functional form of $\gamma_3 (z)$ is
\begin{equation}
\gamma_3 (z) =\gamma_{3}^{(Cd)}(\theta (-d/2 - z)+\theta (z-d/2))
+\gamma_{3}^{(Hg)}(\theta (z+ d/2) -\theta(z-d/2)).
\end{equation}\noindent The $z$-derivative acting on this function
produces $\delta$-function terms that contribute a term
proportional to $(\gamma_{3}^{(Cd)}-\gamma_{3}^{(Hg)})$ to the
integral which vanish when these material parameters are equal.
The integral then has to be evaluated numerically.

After calculating the matrix-elements we are left with an
effective Hamiltonian parameterized in the following way:
\begin{equation}
{\cal{H}}(k_x,k_y)=\left(\matrix{ \epsilon_k+{\cal{M}}(k)&A
k_{-}&0&0\cr A k_{+}&\epsilon_k-{\cal{M}}(k)&0&0\cr
0&0&\epsilon_k+{\cal{M}}(k)&-A k_{+}\cr 0&0&-A k_{-}&\epsilon_k
-{\cal{M}}(k) }\right)
\end{equation} \noindent where
$\epsilon_k = C-D(k_{x}^2+k_{y}^2),$ ${\cal{M}}(k)=M-B
(k_{x}^2+k_{y}^2),$ $k_{\pm}=k_x\pm i k_y,$ and  $A,B,C,D,M$
depend on the specified quantum-well width. For values of these
parameters at $d=40 \AA$ and $d= 70 \AA$ see Table {\it 1}. This
Hamiltonian is block diagonal and can be written in the
form\begin{equation} {\cal{H}}(k_x,k_y)=\left(\matrix{H(k) & 0 \cr
0 & H^{\ast}(-k)}\right)
\end{equation}\noindent where $H(k)
=\epsilon_k I_{2\times 2} + d^a (k)\sigma^a,$ with $d^1=A
k_x,d_2=A k_y,$ and $d^3={\cal{M}}(k)=M-B(k_{x}^2+k_{y}^2).$
Finally, we define a unitary transformation:
\begin{equation}U=
\left(\matrix{ I_{2\times 2}& 0\cr
0&-\sigma^{z}}\right)\end{equation} \noindent and take
$U^{\dagger}H_{4\times 4}U$ which reverses the sign of the linear
$k$ terms in the lower block and puts it ${\cal{H}}$ into the form
\begin{equation}
{\cal{H}}(k_x,k_y)=\left(\matrix{H(k) & 0 \cr 0 &
H^{\ast}(k)}\right).\end{equation} \noindent $U$ does not affect
the $z$-direction of spin and simply rotates the $x$ and $y$ axes
in the lower block of the Hamiltonian by $\pi.$  The energy
dispersions for these bands are given in Fig. {\it 1} of the
supporting online material for several values of $d.$
\begin{table}
\begin{center}
\begin{tabular}{|c|c|c|c|c|c|}
\hline
$d$ ($\AA$) & $A(eV)$ & $B(eV)$ & $C(eV)$ & $D(eV)$ & $M(eV)$\\
\hline 58 &-3.62 & -18.0 &-0.0180 & -0.594 &0.00922\\
\hline 70 &-3.42 & -16.9 &-0.0263 & 0.514 &-0.00686\\
\hline
\end{tabular}
\caption{Parameters for Hg$_{0.32}$Cd$_{0.68}$Te/HgTe quantum
wells.}  \label{HamiltonianParameters}
\end{center}
\end{table}

\begin{figure*}
        \includegraphics[scale=0.55]{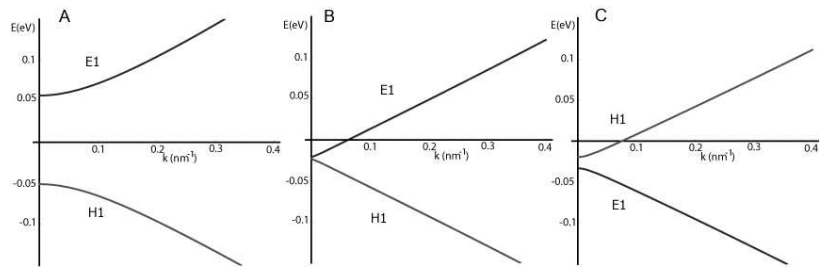}
\caption{ Dispersion relations for the $E1$ and $H1$ subbands
for(A) $d=40 \AA$ (B) $d=63.5\AA$ (C) $d=70\AA.$ }\label{firstfig}
  \end{figure*}

\newpage

\subsection*{Supporting References and Notes}
\begin{itemize}
\item[S1.] \textrm{E.G. Novik} \emph{et. al.}, {\it Phys. Rev.
B\/} {\bf 72}, 035321
  (2005). \item[S2.] A.~Pfeuffer-Jeschke, Ph.D. Thesis, University of Wurzburg, Germany, 2000.
\end{itemize}

\end{document}